\documentclass[oneside]{elsart}
\usepackage[]{fontenc}
\usepackage[latin1]{inputenc}
\usepackage{calc}
\usepackage{graphicx}
\IfFileExists{url.sty}{\usepackage{url}}
                      {\newcommand{\url}{\texttt}}

\makeatletter

\providecommand{\tabularnewline}{\\}

\usepackage{html}
\def\apj{ApJ}                 
\def\mnras{MNRAS}             

\makeatother
\begin{document}
\begin{frontmatter}
\title{Window To The Stars}
\author{Robert G. Izzard\corauthref{cor}},\corauth[cor]{Corresponding author.} ~
\author{Evert Glebbeek}
\address{Sterrenkundig Instituut, Universiteit Utrecht, P.O. Box 80000, NL-3508 TA Utrecht, The Netherlands.
Email: R.G.Izzard@phys.uu.nl}
\date{39th of Dismember, 1791}

\begin{frontmatter}
\begin{abstract}
We present \emph{\normalsize Window To The Stars}{\normalsize , a
graphical user interface to the popular} \emph{\normalsize TWIN} {\normalsize single/binary
stellar evolution code, for novices, students and professional astrophysicists.
It removes the drudgery associated with the traditional approach to
running the code, while maintaining the power, output quality and
flexibility a modern stellar evolutionist requires. It is currently
being used for cutting edge research and interactive teaching.}{\normalsize \par}
\end{abstract}
\end{frontmatter}
\end{frontmatter}

\section{Introduction}

We have developed a graphical user interface, \emph{Window~To~The~Stars}
(\emph{WTTS)}, to Peter Eggleton's popular \emph{TWIN} single and
binary stellar evolution code. We combine a simple interface for input
physics with a comprehensive package of two- and three-dimensional
graph plotters for immediate analysis of results and production of
publication quality plots. \emph{WTTS} is currently being used for
both research and teaching at the Universities of Utrecht and Nijmegen.
It is written in \emph{Perl} and uses the \emph{GTK2} library, both
of which are freely available and portable to all common computing
platforms. We have tested it on Linux and Mac OS X and are working
on a Windows version. \emph{Window~To~The~Stars} can be downloaded
from its homepage at \url{http://www.astro.uu.nl/~izzard/window/ }
together with installation instructions, the \emph{TWIN} code and
some screenshots.

We describe briefly the TWIN code (section \ref{sec:The-TWIN-code})
then the \emph{Window~To~The~Stars} front-end (section \ref{sec:Window-To-The-Stars}).
We conclude with a discussion of current use and future plans in section
\ref{sec:Conclusions}.

\section{The \emph{TWIN} code}

\label{sec:The-TWIN-code}The stellar evolution code used by \emph{WTTS}
is a modified version of the \emph{TWIN} stellar evolution code originally
developed by Peter Eggleton  \cite{1971MNRAS.151..351E,2002ApJ...575..461E}.
It uses an adaptive non-Lagrangian mesh which allows stars to be evolved
with only 200 meshpoints in a few minutes on a normal desktop computer. 

The code has three modes of operation: one to calculate the evolution
of single stars and two different ways for binaries, \emph{TWIN} and
non-\emph{TWIN}. In \emph{TWIN} mode, the two stars are solved for
simultaneously by inverting a single matrix, while in non-\emph{TWIN}
mode the code evolves both stars individually and alternates between
them at regular intervals.

\section{\emph{Window To The Stars} User Interface}

\label{sec:Window-To-The-Stars}
\begin{figure}
\begin{tabular}{cc}
\includegraphics[scale=0.17]{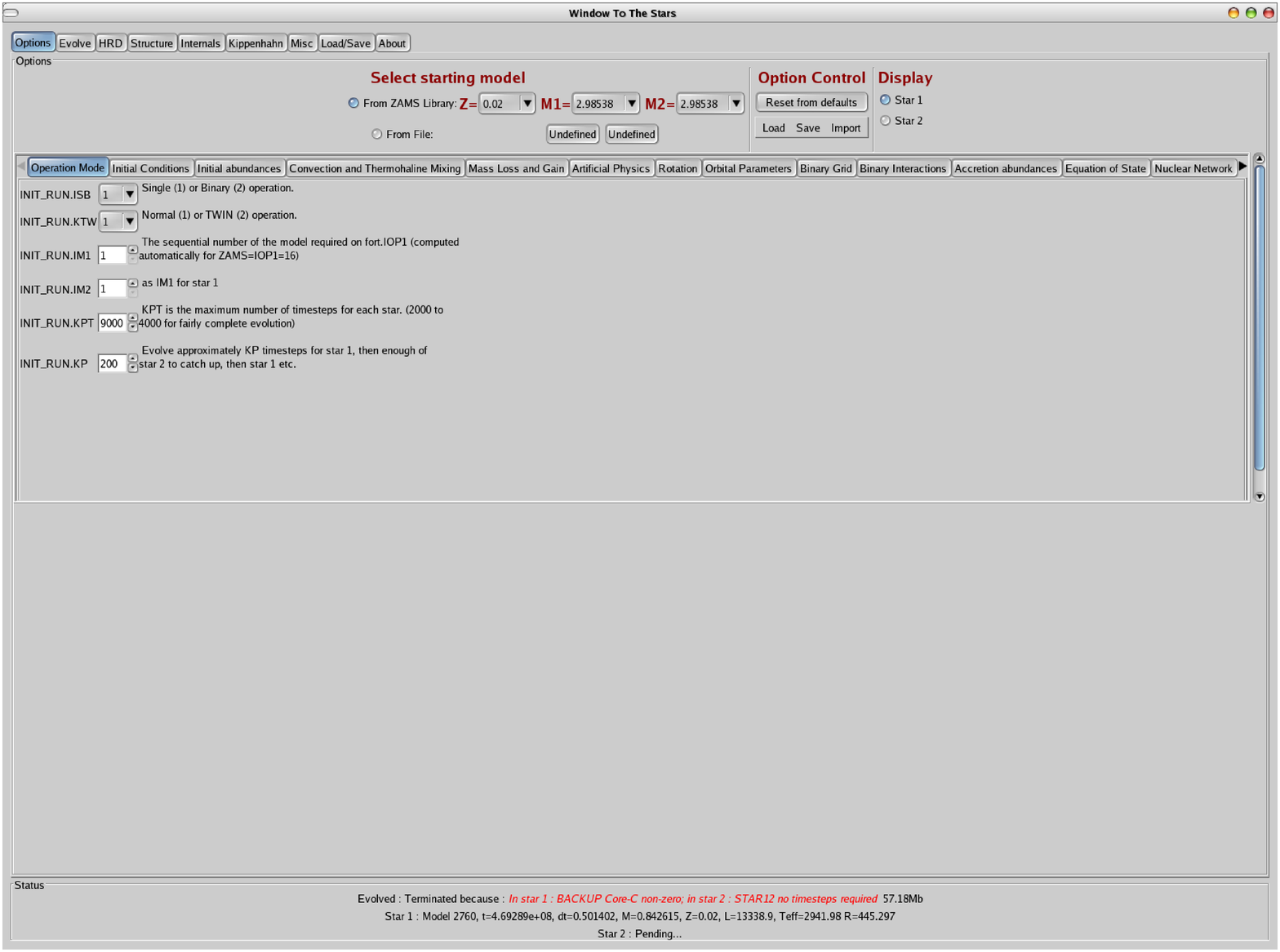}&
\includegraphics[scale=0.17]{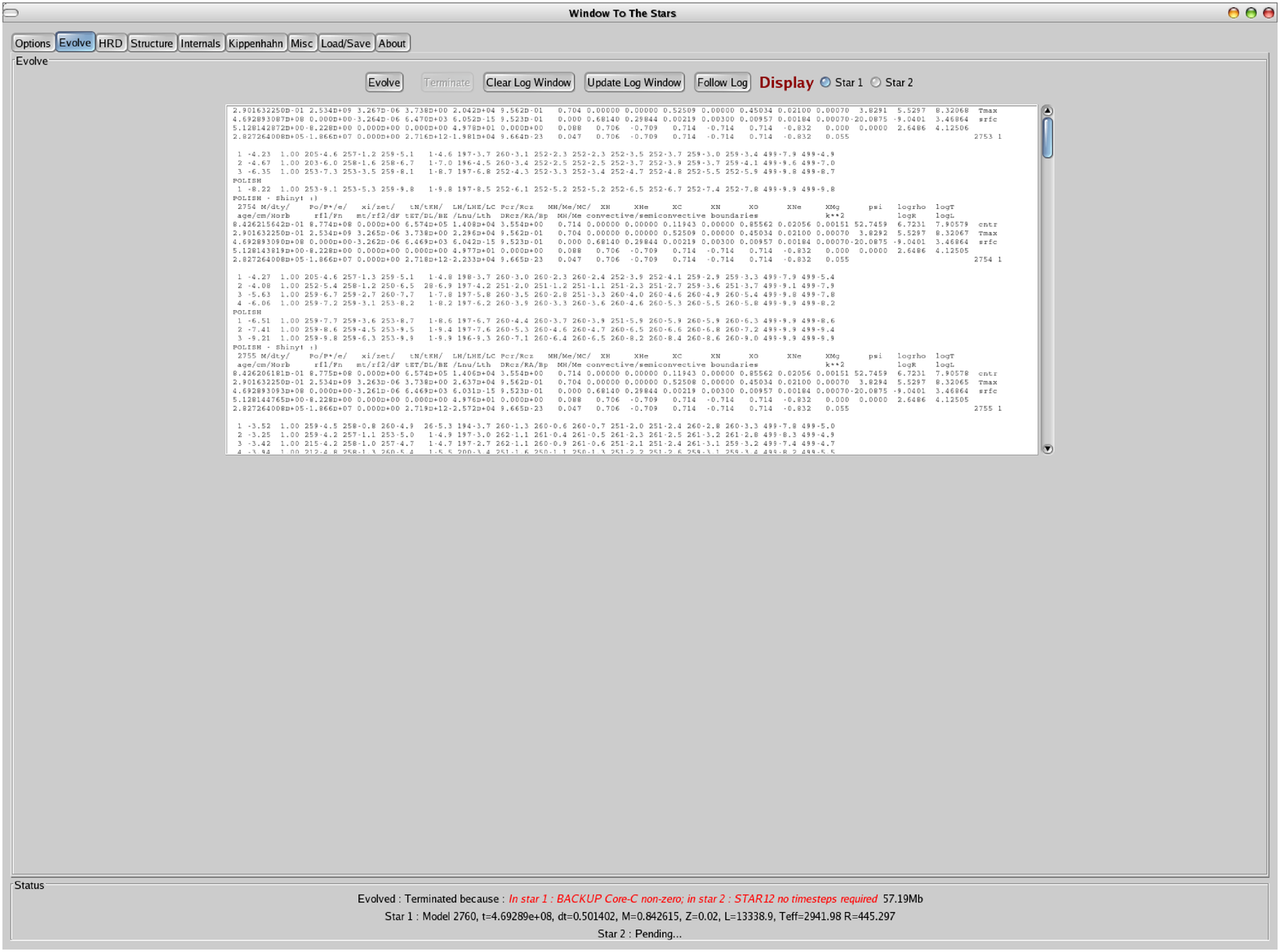}\tabularnewline
\begin{minipage}[t][1\totalheight]{0.45\columnwidth}%
The \emph{Options} tab where physical input is chosen.%
\end{minipage}%
&
\begin{minipage}[t][1\totalheight]{0.45\columnwidth}%
The \emph{Evolve} tab follows the \emph{TWIN} log file.%
\end{minipage}%
\tabularnewline
\includegraphics[scale=0.17]{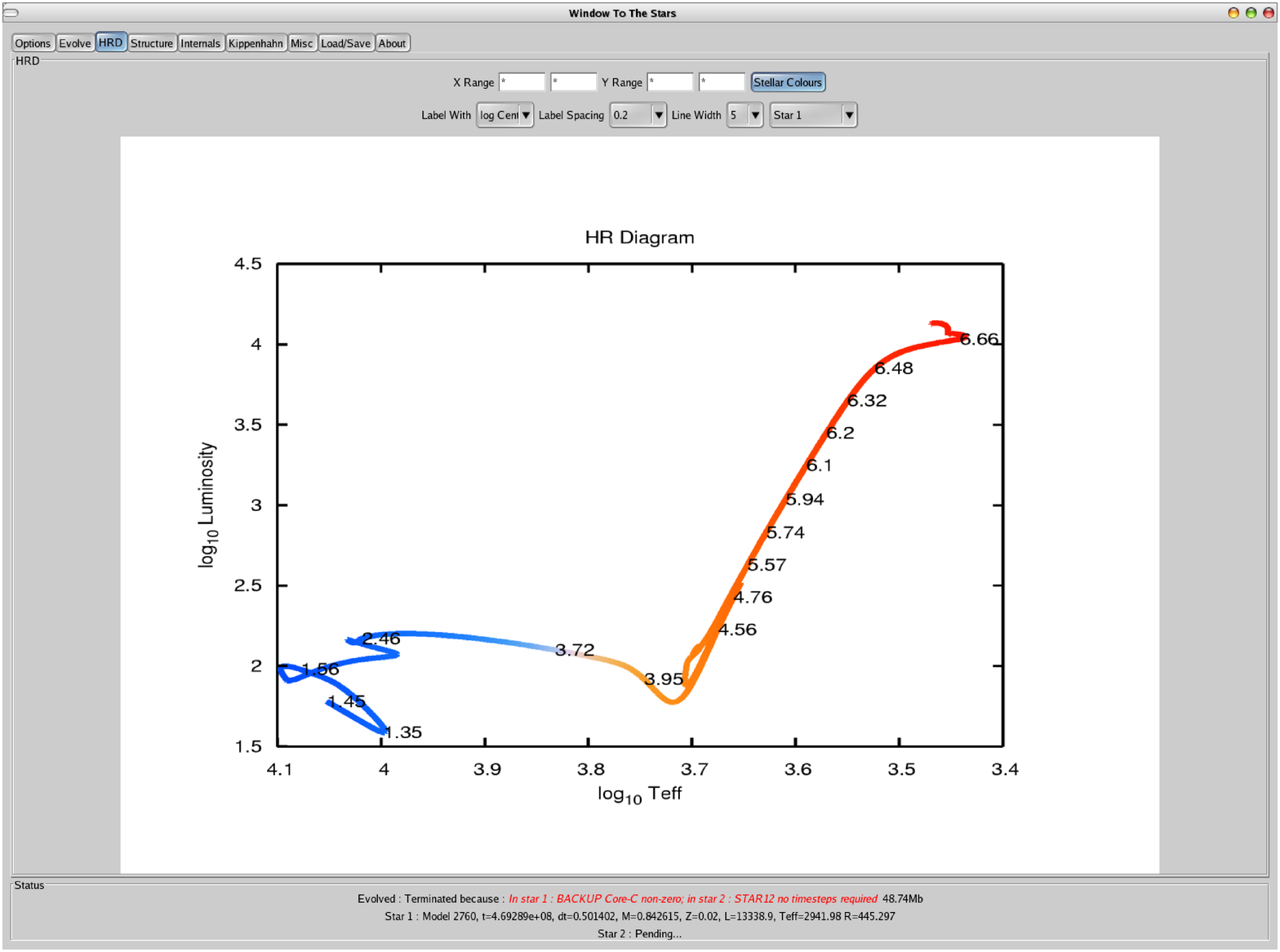}&
\includegraphics[scale=0.17]{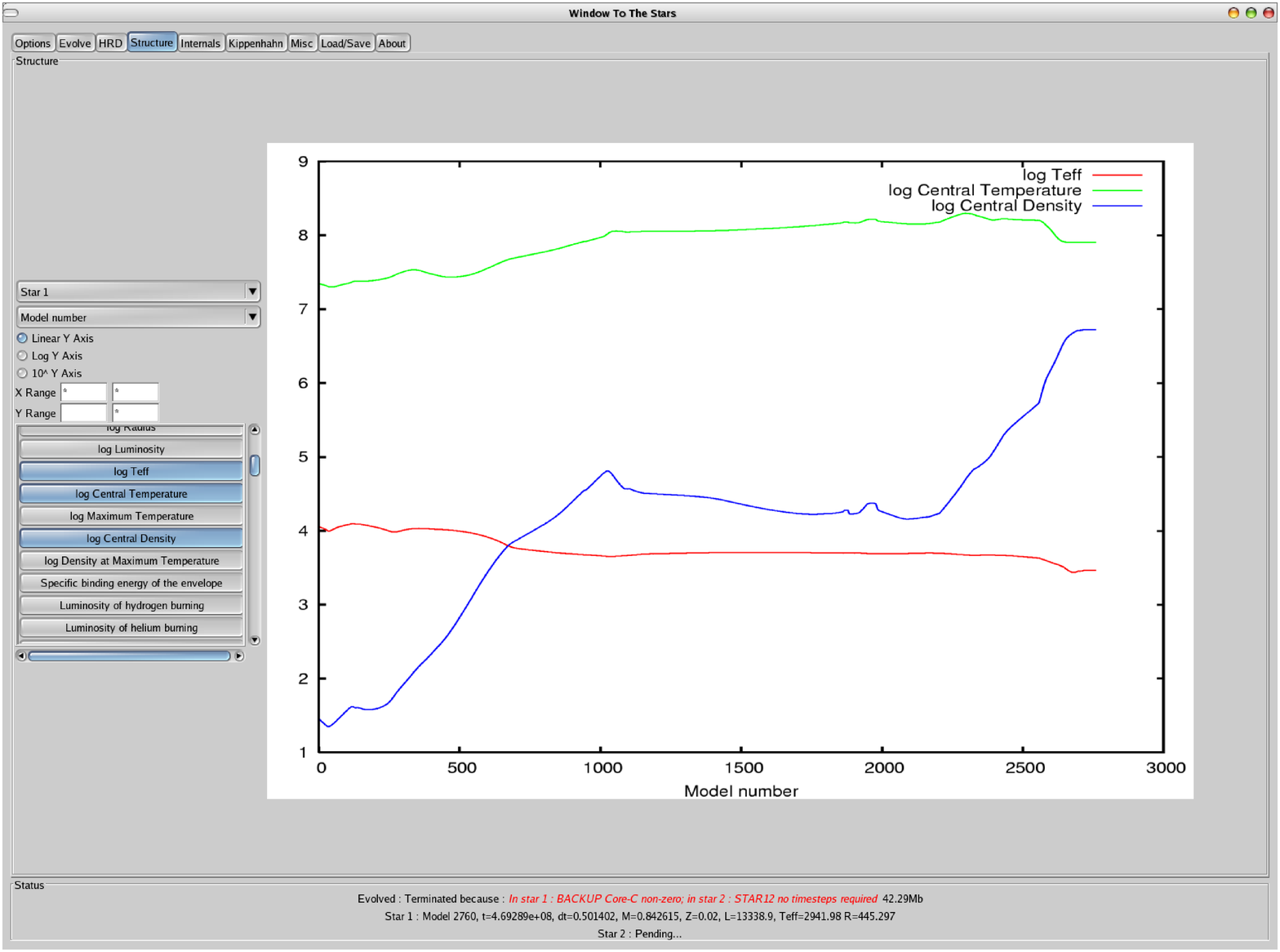}\tabularnewline
\begin{minipage}[t][1\totalheight]{0.45\columnwidth}%
The \emph{HRD} tab, labelled by log central density.%
\end{minipage}%
&
\begin{minipage}[t][1\totalheight]{0.45\columnwidth}%
The \emph{Structure} tab showing surface and central temperature,
and central density.%
\end{minipage}%
\tabularnewline
\includegraphics[scale=0.17]{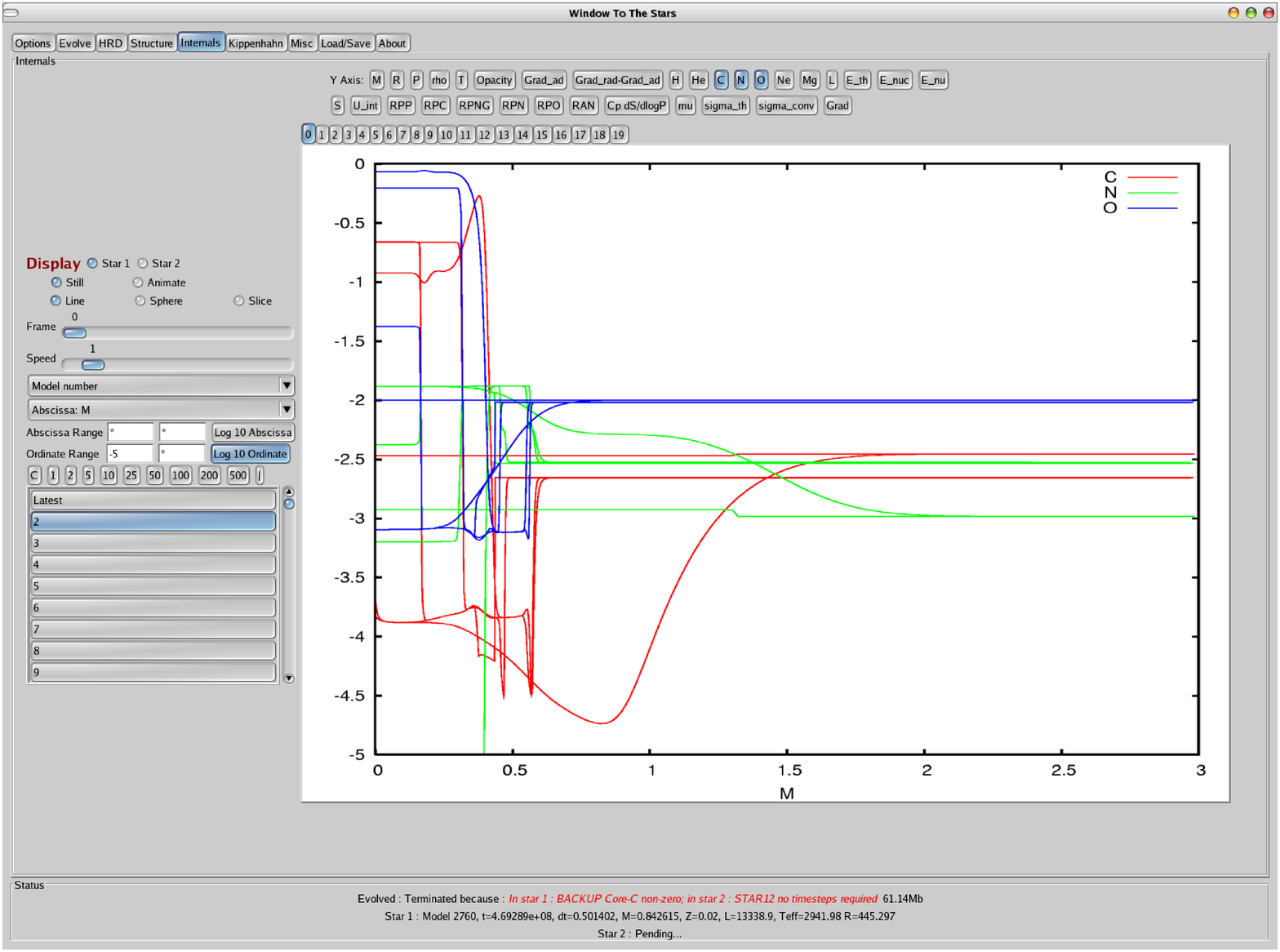}&
\includegraphics[scale=0.17]{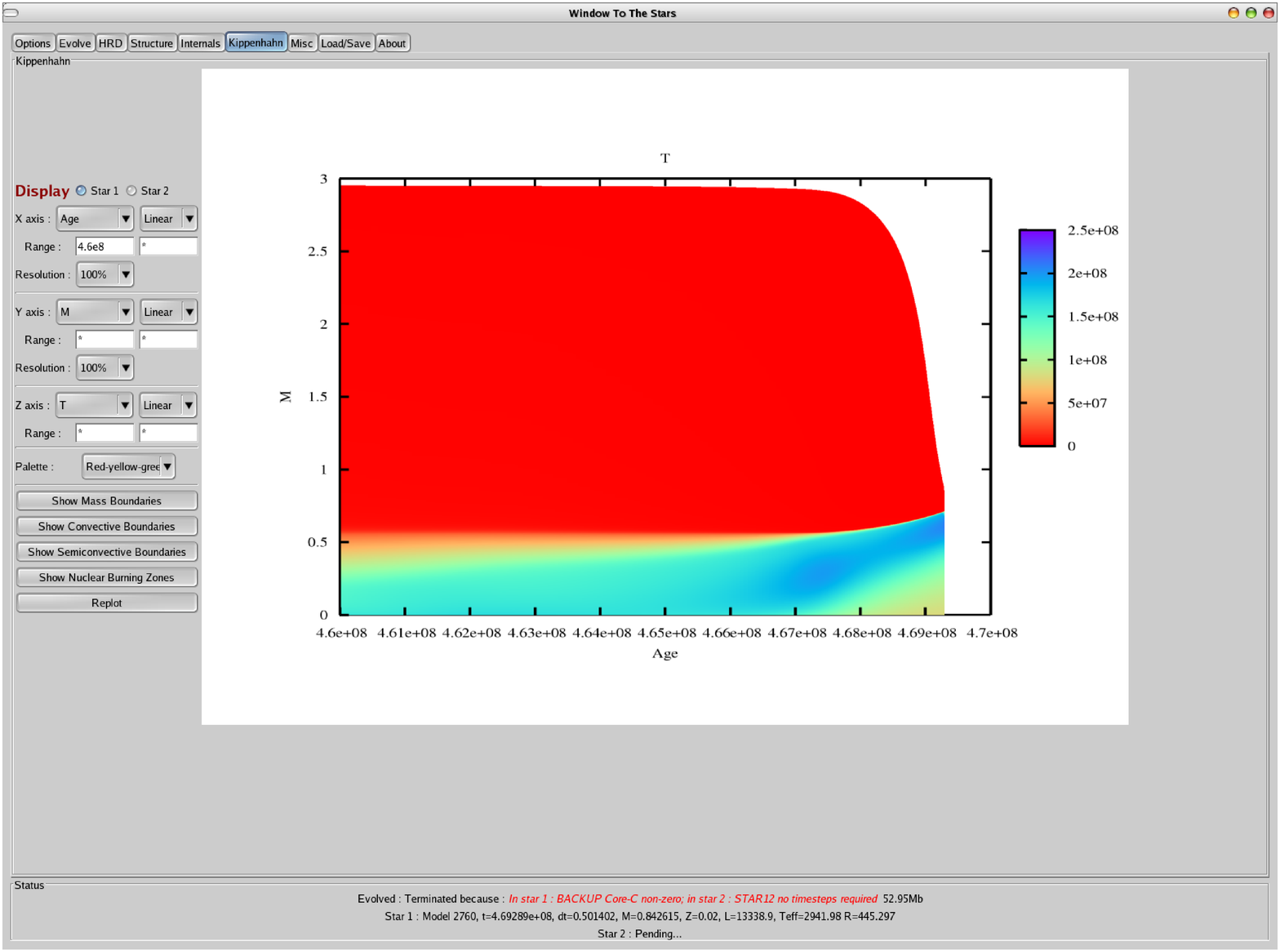}\tabularnewline
\begin{minipage}[t][1\totalheight]{0.45\columnwidth}%
The \emph{Internals} tab showing CNO abundances for one in every 500
models.%
\end{minipage}%
&
\begin{minipage}[t][1\totalheight]{0.45\columnwidth}%
The \emph{Kippenhahn} diagram tab showing temperature as a function
of mass coordinate and age.%
\end{minipage}%
\tabularnewline
\end{tabular}

\caption{\label{fig:Example-screenshots}Example screenshots of the various
tabbed windows showing the evolution of a $3\mathrm{\, M_{\odot}}$
solar metallicity star (a total of 2760 stellar models).}
\end{figure}

\emph{WTTS} is laid out in a logical
progression of tabbed windows which guide the user through the choice
of physics, running of the stellar evolution code and analysis of
the resulting models.  We shall describe each in turn and corresponding
screenshots from the evolution of a $3\mathrm{\, M_{\odot}}$ star
can be found in Fig. \ref{fig:Example-screenshots}. Further examples
and screenshots can be found on the \emph{WTTS} website at \url{http://www.astro.uu.nl/~izzard/window/screenshots/ }.

\begin{description}
\item [{Options}] The first steps in creating a series of stellar evolution
models are the choices of input parameters and physics. We provide
a selection of initial masses from a library of zero-age main sequence
(ZAMS) stellar models with a range of metallicities and the ability
to load pre-calculated models as a starting model. The physics and
initial conditions for stellar evolution are then selected in a series
of tabbed menus such that every aspect of the stellar evolution run
can be controlled. We have added the ability to load or save a complete
set of \emph{WTTS} options, and also to import the original initialisation
(\texttt{init.dat}) files from the \emph{TWIN} code distribution.
\item [{Evolve}] The bulk of the stellar evolution is done here. There
are buttons to start and stop the evolutionary sequence. The \emph{TWIN}
log file is also shown and updated automatically during the evolutionary
run.
\item [{HRD}] The Hertzsprung-Russell diagram (HRD), which shows $\log L/\mbox{L}_{\odot}$
vs $\log T_{\mathrm{eff}}/\mbox{K}$, is displayed for each star individually
or for both stars, and updates as the evolution progresses. The tracks
can be labelled with a third variable, such as central temperature
or core mass, and coloured according to the effective temperature.
\item [{Structure}] In this tab structural variables such as age, model
number, total mass, core mass, luminosity, radius and abundances are
plotted against one another. The plot automatically updates throughout
the evolution and many variables can be plotted at once, with the
option of logging the axes and choosing data ranges.
\item [{Internal}] It is often desirable to examine the internal details
of each stellar model individually or plot the results from a few
models in one graph. Successive model information can be overlaid
or turned into an animated sequence. Each model can be saved and later
used as a starting model for a new evolutionary run (see the \emph{Options}
tab description).
\item [{Kippenhahn}] The traditional Kippenhahn plot shows convective regions
as a function of mass coordinate and time. We expand on this idea
by plotting a mapped 3D surface of any of the stellar evolutionary
variables as a function of any two others. In the case of mass-coordinate
used as an ordinate, the core mass(es), convective boundaries and
nuclear burning zones can be over-plotted. 
\item [{Miscellaneous}] Other options can be changed here, such as image
size, HRD colour contrast and hue, font type and size, plot type (postscript
or PNG) etc.
\end{description}
In addition to the main tab window, a status bar is visible at all
times, informing the user of the current status (e.g. {}``Evolving''
or {}``Stopped'') and the physical state of the primary and secondary
star.

All images can be saved as either PNG, suitable for web-based projects,
or Postscript for publications and talks. Animations are saved as
animated GIFs, also compatible with web browsers.

\subsection{Installation}

We provide detailed installation instructions on our website \url{http://www.astro.uu.nl/~izzard/window/INSTALL}
as well as an automated installation script which should work with
most incarnations of Linux and possibly Mac OS X.

\emph{WTTS} depends on some free software packages: Perl, GTK2, Perl-GTK,
Gnuplot, Imagemagick, Ghostscript and a number of Perl modules (see
\url{http://www.perl.com/}, \url{http://www.gtk.org/}, \url{http://www.imagemagick.org},
\url{http://www.cs.wisc.edu/~ghost/}, \url{http://www.gnuplot.info/}
and CPAN, \url{http://www.cpan.org}, for Perl modules). All these
packages are available for, or are installed by default on, a modern
Linux system; we have also had success running \emph{WTTS} on a Mac
running OS X. A Fortran compiler is required to build \emph{TWIN}:
it is compatible with the GNU \emph{g95} compiler (\url{http://g95.sourceforge.net/}),
which is free and portable, as well as Intel's not really free compiler.
We have also made some binary executables of \emph{WTTS} and \emph{TWIN}
available on our website, for Linux \texttt{i386} (32-bit) and \texttt{x86\_64}
(64-bit) systems.

\section{Conclusions and Future Plans}

\emph{\label{sec:Conclusions}Window To The Stars} provides everybody,
from novices to experts, with a powerful yet uncomplicated stellar
evolution tool. Its graphical interface circumvents the need for manual
hacking of tables of cryptic numbers. It is capable of evolving stars
and analysing the results, with output of publication quality. The
focus is on physics, not computing. While \emph{WTTS} has reached
an advanced level, we plan numerous improvements, most notably the
ability to transparently use different stellar evolution codes with
the same graphical user interface.

Paul Groot of the University of Nijmegen is the first person to use
\emph{WTTS} in a stellar evolution course, and he describes his experience
so far as {}``extremely positive''. The 2007 stellar evolution course
at Utrecht University will include exercises based on \emph{WTTS}.
We already use \emph{WTTS} to analyze our work on stellar accretion
and mergers.

\bibliographystyle{plain}

\begin{thebibliography}{1}

\bibitem{1971MNRAS.151..351E}
P.~P. {Eggleton}.
\newblock {The evolution of low mass stars}.
\newblock {\em \mnras}, 151:351, 1971.

\bibitem{2002ApJ...575..461E}
P.~P. {Eggleton} and L.~{Kiseleva-Eggleton}.
\newblock {The Evolution of Cool Algols}.
\newblock {\em \apj}, 575:461--473, August 2002.

\end{thebibliography}

\begin{ack}
RGI would like to thank everyone who contributed to this project,
especially his in-laws for boring him into starting the project and
Cecilia and Stefan for feeding him during a stay in Heidelberg, where
\emph{WTTS} grew up. RGI and EG are supported by the Nederlandse Organisatie
voor Wetenschappelijk Onderzoek (NWO).
\end{ack}

\end{document}